# Stability and Sensitivity Analysis of Multi-Vendor, Multi-Terminal HVDC Systems

Yicheng Liao
School of Electrical Engineering and Computer Science
KTH Royal Institute of Technology, Stockholm, Sweden
ycliao@ieee.org

Heng Wu, Xiongfei Wang
AAU Energy
Aalborg University, Aalborg, Denmark
{hew, xwa}@energy.aau.dk

Mario Ndreko, Robert Dimitrovski, Wilhelm Winter
Electrical System Design
TenneT TSO GmbH, Bayreuth, Germany
{mario.ndreko, robert.dimitrovski, wilhelm.winter}@tennet.eu

*Abstract*—The stability of multi-vendor, multi-terminal HVDC systems can be analyzed in frequency domain by black-box impedance models using the generalized Nyquist stability criterion. Based on the impedance stability analysis, a multi-level sensitivity analysis approach using frequency-domain sensitivity functions is proposed to identify the root cause of potential instability. Case studies on a four-terminal HVDC system are carried out for stability and sensitivity analysis based on the impedance measurement in PSCAD. The analysis results are finally validated by electromagnetic transient simulations.

## I. Introduction

Power electronic based converters play important roles in high-voltage direct-current (HVDC) transmission systems for efficient and flexible power supply. However, due to the fast dynamics of converter's control, the interactions among power converters and power grids can lead to oscillations in a wide frequency range [1], [2], which could jeopardize the power quality and may even threaten the stable operation of the power system.

In order to enable the development of future multi-terminal HVDC systems, the interoperability of multi-vendor projects should be ensured in the planning and design phase [3], [4]. It is usually the responsibility of transmission system operators (TSOs) or if applicable the owner of such HVDC systems to conduct together with the HVDC suppliers a system-level stability assessment to prove that HVDC converters can be connected at the DC side without interoperability risks of its control systems. To assess the control-induced stability of such a multi-vendor multi-terminal HVDC system, a common way is to use electromagnetic transient (EMT) simulations (either on offline or real time platforms) [2], [5]. Although, the EMT simulation is the safest way to prove the stability of the overall system as it encompassed detailed modelling up to component level, it cannot provide sufficient insights to the instability mechanism taking placed when adverse interactions are observed. As a consequence, a lot of research works have been carried out in recent years on the dynamical modeling and analysis of HVDC systems [6]-[9]. Among the existing studies, the impedance-based approach [6], [9] becomes more and more important since it can model the converters at the connection points where the study is performed as a black-box system. The dynamic behaviors of such converters can be characterized by impedance models at their points of connection through frequency-scan measurement performed on an EMT simulation platform, and the system stability can be further assessed by the frequency-domain tools. This approach also provides some insights into the instability mechanism, that is the negative resistance caused by converter controls can cause instability issues when interacting with grid impedances [10].

However, when multiple converters are interconnected to each other through complex DC network, the stability becomes more complicated, since multiple converters may introduce negative resistance which can interact with each other. A few recent publications have studied the stability and sensitivity of multi-converter based systems using impedance models [11]-[13]. There are two major approaches. One is based on the pole-residue relationship of transfer functions, where the sensitivity of the system oscillation modes with respect to impedance models is derived to characterize the impacts of nodal voltages or branch currents on the system stability [11]. Furthermore, the parameter sensitivity can be deduced based on the chain rule if the sensitivity of impedance models with respect to any parameter is known [12]. However, this method requires pre-known *s*-domain transfer functions for the pole and residue calculation, which are usually not accessible for black-box systems. The other approach is based on the frequency-domain sensitivity functions, which is adopted in [13] to study the sensitivity analysis of a multi-terminal HVDC system. The sensitivity functions of the determinant of the system return-difference matrix with respect to impedance models are derived to characterize the converters' impacts on the system stability. However, this approach relies on the assumption that the converter admittance models do not have right-half-plane poles. Thus, additional model modifications are needed when the assumption does not hold, which complicates the analysis. Furthermore, the DC network is only regarded as a single-port network for each converter station, which does not apply to the station including both positive and negative poles at the DC side.

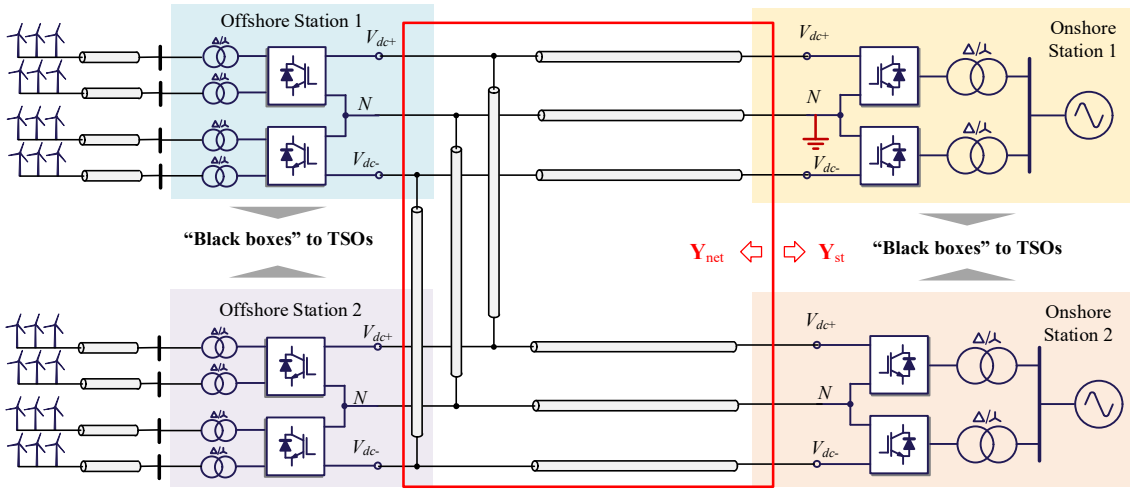

Fig. 1 Four-terminal MMC-HVDC system under study

To address the aforementioned challenges, a multi-level sensitivity analysis approach is proposed to study the stability and sensitivity of multi-vendor, multi-terminal HVDC systems. Differing from [13], the impedance-based stability is analyzed by the generalized Nyquist stability criterion (GNSC) applied to the system return ratio. Then, multi-level sensitivity functions of the return-ratio eigenlocus are derived, which are combined with the stability analysis results to identify the root cause of potential instability. The proposed approach can be implemented based on black-box impedance models and does not require the system to be minimum-phase, which benefits to the system-level studies for TSOs. In the rest of the paper, Section II introduces the system under study and reviews the basic impedance-based stability criterion. Section III proposes the multi-level sensitivity analysis approach. Section IV provides the case studies based on measured impedance models. Conclusions are drawn in Section V.

## II. SYSTEM DESCRIPTION AND IMPEDANCE-BASED STABILTY CRITERION

### A. System Description

Fig. 1 shows an offshore four-terminal HVDC system under study, which is based on modular multilevel converters (MMCs). Two onshore stations and two offshore stations are connected through HVDC cables. The onshore stations are controlled in the grid-following mode, while the offshore stations are controlled in the grid-forming mode (applying V/f control). It is assumed in this paper that the MMC HVDC stations are manufactured by different vendors, thus, they can only be regarded as black-box models to the TSOs for the system-level study.

The impedance models are used to characterize the system dynamic behaviors, which can be obtained through impedance measurement obtained from EMT simulation platforms [14]. To conduct the impedance-based stability analysis, the entire system is partitioned into one active subsystem and one passive subsystem by multiple points of connection (PoCs), as shown by the red box in Fig. 1. The active subsystem includes all the MMC stations, whose admittance model can be represented by $\mathbf{Y}_{st}$. The passive subsystem is composed by the DC network, whose admittance model is denoted by $\mathbf{Y}_{net}$.

Through the subsystem partition, the entire system can be treated as a multi-port cascaded system, as shown in Fig. 2. Each MMC station has a positive pole and a negative pole, whose dynamics can be represented by two DC impedances. Thus, each MMC station is regarded as a three-port system at the PoCs. Any DC cable among two MMC stations can be represented by a six-port system. The four-terminal HVDC system is thus represented as a twelve-port cascaded system.

The voltages at the PoCs are defined by $V_k$. The currents flowing from the MMCs to the DC network are defined as $I_k$. Then, the MMC station admittance can be derived as

$$\mathbf{Y}_{st} = \begin{bmatrix} \mathbf{Y}_{on1} & & & \\ & \mathbf{Y}_{off1} & & \\ & & \mathbf{Y}_{off2} & \\ & & & \mathbf{Y}_{on2} \end{bmatrix} \quad (1)$$

where $\mathbf{Y}_{on1}$, $\mathbf{Y}_{off1}$, $\mathbf{Y}_{off2}$, $\mathbf{Y}_{on2}$ represent the admittance matrices of the Onshore Station 1, Offshore Station 1, Offshore Station 2, and Onshore Station 2, respectively.

Without losing generality, each MMC station admittance can be denoted by its positive-pole and negative-pole impedances. Taking the Onshore Station 1 as an example, its admittance is deduced as

$$\mathbf{Y}_{on1} = \begin{bmatrix} \dfrac{1}{Z_{on\_p1}} & -\dfrac{1}{Z_{on\_p1}} & \\ -\dfrac{1}{Z_{on\_p1}} & \dfrac{1}{Z_{on\_p1}} + \dfrac{1}{Z_{on\_n1}} & -\dfrac{1}{Z_{on\_n1}} \\ & -\dfrac{1}{Z_{on\_n1}} & \dfrac{1}{Z_{on\_n1}} \end{bmatrix} \quad (2)$$

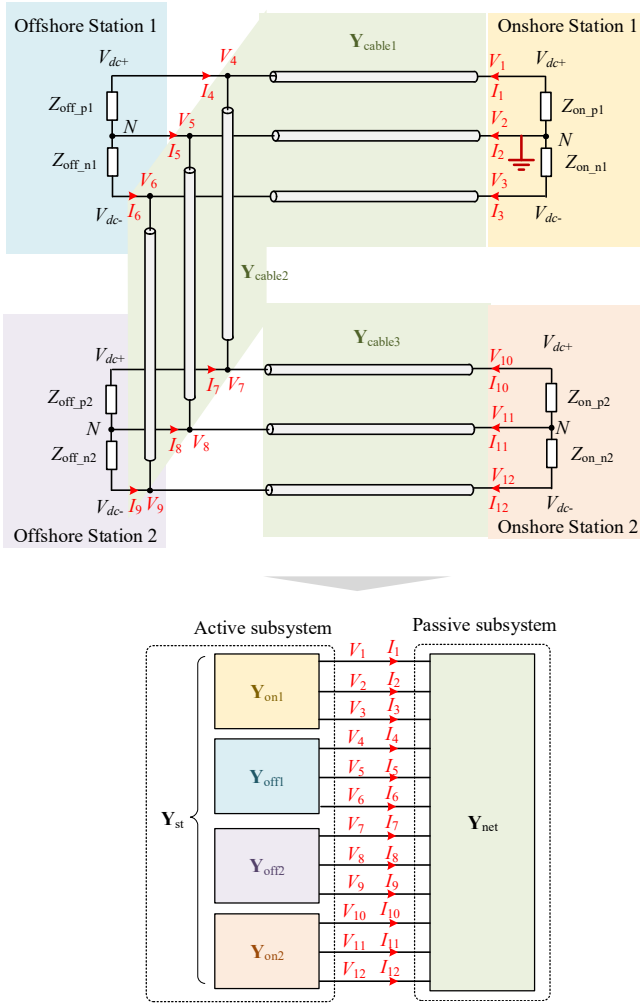

Fig. 2 Subsystem partition for admittance representation

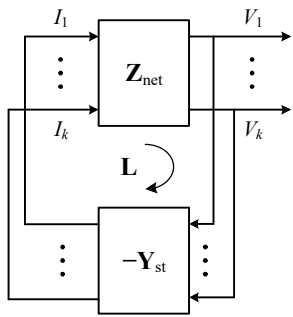

Fig. 3 Four-terminal MMC-HVDC system under study

where $Z_{on\_p1}$ is the positive-pole impedance and $Z_{on\_p1}$ is the negative-pole impedance.

It is assumed that each cable admittance is represented by $\mathbf{Y}_{cablek}$, which should be a six-by-six matrix according to the number or PoCs. Then, based on the cable configurations, the DC network admittance can be deduced as

$$\mathbf{Y}_{net} = \begin{bmatrix} \square & \square & \mathbf{O} & \mathbf{O} \\ \square & \square & \mathbf{O} & \mathbf{O} \\ \mathbf{O} & \mathbf{O} & \mathbf{O} & \mathbf{O} \\ \mathbf{O} & \mathbf{O} & \mathbf{O} & \mathbf{O} \end{bmatrix} + \begin{bmatrix} \mathbf{O} & \mathbf{O} & \mathbf{O} & \mathbf{O} \\ \mathbf{O} & \square & \square & \mathbf{O} \\ \mathbf{O} & \square & \square & \mathbf{O} \\ \mathbf{O} & \mathbf{O} & \mathbf{O} & \mathbf{O} \end{bmatrix} + \begin{bmatrix} \mathbf{O} & \mathbf{O} & \mathbf{O} & \mathbf{O} \\ \mathbf{O} & \mathbf{O} & \mathbf{O} & \mathbf{O} \\ \mathbf{O} & \mathbf{O} & \square & \square \\ \mathbf{O} & \mathbf{O} & \square & \square \end{bmatrix} \quad (3)$$

where $\mathbf{O}$ or $\square$ represents a three-by-three block matrix. $\mathbf{O}$ denotes a zero matrix and $\square$ denotes a non-zero matrix. It can be seen that $\mathbf{Y}_{cable1}$ is associated to the Ports 1-6, since it connects the Onshore Station 1 and the Offshore Station 1. $\mathbf{Y}_{cable2}$ is associated to the Ports 4-9, since it connects the Offshore Station 1 and the Offshore Station 2. $\mathbf{Y}_{cable3}$ is associated to the Ports 7-12, since it connects the Offshore Station 2 and the Onshore Station 2.

*B. Impedance-Based Stability Criterion*

The interconnected system in Fig. 2 can be formulated by a multivariable feedback control system, with the loop gain representation shown in Fig. 3. It is noted that only $\mathbf{Y}_{net}$ can be inversed as $\mathbf{Z}_{net}$, since Eq. (2) has the rank of two, implying that $\mathbf{Y}_{st}$ cannot be inversed.

Then, the system return ratio is defined as

$$\mathbf{L}(s) = \mathbf{Y}_{st}(s)\mathbf{Z}_{net}(s) \quad (4)$$

and the system return difference is defined as

$$\mathbf{F}(s) = \mathbf{E} + \mathbf{L}(s) \quad (5)$$

where $\mathbf{E}$ is an identity matrix.

The stability of a multivariable feedback system can be analyzed by the GNSC [15], which in general includes three steps:

- Step 1: Identify the number of open-loop right-half-plane (RHP) poles ($P$) by the poles of $\det(\mathbf{F}(s))$.
- Step 2: Count the total number of anticlockwise encirclements ($N$) around (−1, 0) on Nyquist diagrams of $\lambda_i(s)$, which are the eigenvalues of $\mathbf{L}(s)$.
- Step 3: Check the stability by judging if $P=N$.

It can be seen that the system stability is influenced significantly by the eigenloci of $\mathbf{L}(s)$, thus, the system sensitivity can be analyzed in the frequency domain based on sensitivity functions of $\lambda_i(s)$.

### III. MULTI-LEVEL SENSITIVITY ANALYSIS

A multi-level sensitivity analysis approach is proposed in this section based on the frequency-domain sensitivity theory [16]. According to the generalized NSC, the trajectory of $\lambda_i(s)$ affects the stability significantly. Thus, the sensitivity analysis can be conducted by deriving sensitivity functions of $\lambda_i(s)$ with respect to any other transfer function $G(s)$ of interest, i.e.,

$$S_G^{\lambda_i}(s) = \partial\lambda_i(s)/\partial G(s) \quad (6)$$

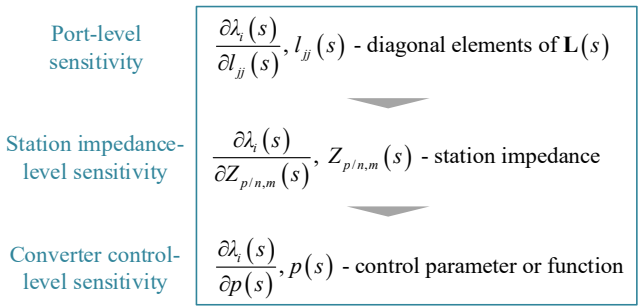

Fig. 4 Multi-level sensitivity analysis approach

The frequency-domain sensitivity function in (6) allows for doing sensitivity analysis at different levels, depending on which $G(s)$ is chosen. A multi-level sensitivity analysis approach is developed in Fig. 4. The sensitivity can be analyzed at the port level, station impedance level, and the converter control level.

### A. Port-Level Sensitivity (Port Participation)

The port-level sensitivity quantifies the impacts of different port voltages/currents on the eigenlocus $\lambda_i(s)$.

According to Fig. 3, there exists

$$[\mathbf{E} + \mathbf{L}(s)][I_1 \cdots I_k]^T = \mathbf{0} \quad (7)$$

Thus, the return ratio $\mathbf{L}(s) = \mathbf{Y}_{st}(s)\mathbf{Z}_{net}(s)$ denotes how the port currents affect their owns through the entire feedback loop. Similarly, if $\mathbf{L}(s)$ is defined as $\mathbf{Z}_{net}(s)\mathbf{Y}_{st}(s)$, it denotes how the port voltages affect their owns through the entire feedback loop.

Through the eigenvalue decomposition, $\mathbf{L}(s)$ can be represented by the eigenvalue matrix and the eigenvector matrices, i.e.,

$$\mathbf{L}(s) = \mathbf{V}(s)\mathbf{\Lambda}(s)\mathbf{W}(s)$$

where $\mathbf{V}(s)$ - right eigenvector matrix

$\mathbf{\Lambda}(s)$ - eigenvalue matrix

$\mathbf{W}(s)$ - left eigenvector matrix  (8)

Then, the sensitivity function of $\lambda_i(s)$ with respect to $l_{jk}(s)$ can be derived as

$$S^{\lambda_i}_{l_{jk}}(s) = \frac{\partial \lambda_i(s)}{\partial l_{jk}(s)} = w_{ij}(s)v_{ki}(s) \quad (9)$$

where $\lambda_i(s)$ - $i$-th diagonal element of $\mathbf{\Lambda}(s)$

$l_{jk}(s)$ - $j$-th row, $k$-th column element of $\mathbf{L}(s)$

$w_{ij}(s)$ - $i$-th row, $j$-th column element of $\mathbf{W}(s)$

$v_{ki}(s)$ - $k$-th row, $i$-th column element of $\mathbf{V}(s)$

If it is assumed that $j=k$ in (9), the participation matrix of the port currents into the eigenloci can be defined as

$$\mathbf{P}(s) := \begin{bmatrix} S^{\lambda_1}_{l_{11}}(s) & \cdots & S^{\lambda_i}_{l_{11}}(s) \\ \vdots & \ddots & \vdots \\ S^{\lambda_1}_{l_{jj}}(s) & \cdots & S^{\lambda_i}_{l_{jj}}(s) \end{bmatrix} \quad (10)$$

where $S^{\lambda_i}_{l_{jj}}(s) = \frac{\partial \lambda_i(s)}{\partial l_{jj}(s)} = w_{ij}(s)v_{ji}(s)$

The sensitivity matrix in (10) is named as the participation matrix, since its derivation is similar to the participation factors of state-space analysis [17]. The sum of each row or column is equal to one. However, the implication here is different. Each element in (10) is a participation function in the frequency domain.

### B. Station Impedance-Level Sensitivity

The sensitivity analysis can be done further at the station impedance level, by deriving the sensitivity functions of $\lambda_i(s)$ with respect to the station impedance $Z_{p/n,m}(s)$, where the subscript "$p/n$" denotes the positive-pole or negative-pole impedance, and the subscript "$m$" denotes the $m$-th station.

First, the sensitivity function matrix of $\mathbf{L}(s)$ with respect to $Z_{p/n,m}(s)$ can be derived as

$$\mathbf{S}^{\mathbf{L}}_{Z_{p/n,m}}(s) := \frac{\partial \mathbf{L}(s)}{\partial Z_{p/n,m}(s)} = \frac{\partial \mathbf{Y}_{st}(s)}{\partial Z_{p/n,m}(s)} \mathbf{Z}_{net}(s)$$

$$= \begin{bmatrix} \frac{\partial \mathbf{Y}_1(s)}{\partial Z_{p/n,m}(s)} & & & \\ & \frac{\partial \mathbf{Y}_2(s)}{\partial Z_{p/n,m}(s)} & & \\ & & \ddots & \\ & & & \frac{\partial \mathbf{Y}_m(s)}{\partial Z_{p/n,m}(s)} \end{bmatrix} \mathbf{Z}_{net}(s) \quad (11)$$

where $\mathbf{Y}_m(s)$ represents the $m$-th station admittance matrix at its PoCs, which are $\mathbf{Y}_{on1}$, $\mathbf{Y}_{off1}$, $\mathbf{Y}_{off2}$, $\mathbf{Y}_{on2}$ according to (1).

Then, based on the chain rule, the sensitivity function of $\lambda_i(s)$ with respect to $Z_{p/n,m}(s)$ can be derived as

$$S^{\lambda_i}_{Z_{p/n,m}}(s) = \frac{\partial \lambda_i(s)}{\partial Z_{p/n,m}(s)} = \sum_{j,k} S^{\lambda_i}_{l_{jk}}(s) S^{l_{jk}}_{Z_{p/n,m}}(s)$$

where $S^{l_{jk}}_{Z_{p/n,m}}(s)$ is the $j$-th row, $k$-th column element of $\mathbf{S}^{\mathbf{L}}_{Z_{p/n,m}}(s)$  (12)

To compare the eigenlocus sensitivity functions with respect to different station impedances, the sensitivity function of (12) can be normalized as

$$S^{\lambda_i}_{Z_{p/n,m}}(s) = \frac{Z_{p/n,m}(s)}{\lambda_i(s)} \sum_{j,k} S^{\lambda_i}_{l_{jk}}(s) S^{l_{jk}}_{Z_{p/n,m}}(s) \quad (13)$$

### C. Converter Control-Level Sensitivity

According to the chain rule, the sensitivity can be done further at the converter control level, if the sensitivity function of $Z_{p/n,m}(s)$ with respect to any converter control parameter or function of $p(s)$ is known. Then, the eigenlocus sensitivity is derived as

$$S^{\lambda_i}_p(s) = \frac{\partial \lambda_i(s)}{\partial p(s)} = \frac{\partial \lambda_i(s)}{\partial Z_{p/n,m}(s)} \cdot \frac{\partial Z_{p/n,m}(s)}{\partial p(s)} \quad (14)$$

The converter-level sensitivity may not be available for TSOs, since the converter control details are kept confidential by converter manufacturers. However, the multi-level sensitivity analysis approach shows that the sensitivity functions can be transferred from the system level to the converter level. Thus, it provides potentials for converter manufacturers to use the upper-level black-box sensitivity functions for the bottom-level sensitivity analysis, which can benefit to the converter control design.

### D. Sensitivity Analysis Based on Stability Analysis

Based on the frequency-domain stability analysis, the sensitivity analysis can be conducted following the steps below:

- Step 1: Found the critical eigenloci that affect the stability significantly and identify the critical frequency range where the potential instability may happen.

    The critical eigenloci are the ones close to (−1, 0) on the Nyquist diagram, since their encirclements around (−1, 0) are more likely to be changed with system parameter change, which are thus more sensitive to the system stability. The critical frequency range should be the neighborhood where the Nyquist diagram is close to (−1, 0), which can be identified on the Bode diagram. The critical frequency range is usually around the magnitude crossover frequency of 0 dB or the phase crossover frequency of 180°.

- Step 2: Calculate the critical eigenlocus sensitivity functions and compared their magnitude responses in the critical frequency range.

    At the port level, the port participation analysis can be conducted for the critical eigenloci. In the critical frequency range, the ports with highest participations should take dominant roles in the system stability or instability.

    At the station impedance level, the normalized sensitivity functions of critical eigenloci with respect to different station impedances can be compared. In the critical frequency range, the highest sensitivity magnitude indicates which station takes a dominant role in the system stability or instability.

It is noted that the sensitivity analysis is only necessary when the system is unstable, or close to instability with poor damping, i.e., there exist eigenloci of **L**(s) that are close to (−1, 0).

## IV. CASE STUDIES

Two case studies are carried out for the stability and sensitivity analysis of the system in Fig. 1. The impedance models of the MMC-HVDC stations and the DC cables are all measured in PSCAD by the PSCAD-compatible software toolbox developed in [14]. The stability and sensitivity are then analyzed based on the measured impedance models. The analysis results are compared with the EMT simulations for validation.

### A. Case I – Unstable Case

First, an unstable case is studied. The stability analysis results are provided in Fig. 5. Fig. 5(a) shows the Bode diagram of det(**F**(s)). It can be seen that at all the resonant peaks, the phase is decreasing, indicating that det(**F**(s)) has none RHP poles [18]. Thus, $P$=0. Fig. 5(b) shows the Nyquist diagrams of all the eigenloci of **L**(s) within the positive frequency range. It can be seen that $\lambda_1(s)$ and $\lambda_2(s)$ encircle (−1, 0) clockwise two times in total. If the negative frequency range is considered, the total number of clockwise encirclements is four, i.e., $N$=−4. Thus, the system is unstable.

Fig. 6 provides the EMT simulation results of Case I. Oscillations are observed in the time-domain waveforms. The discrete Fourier analysis of the DC-link voltage in Fig. 6(b) indicates an oscillation frequency of 630 Hz, which closely agrees with the phase crossover frequency of 180° shown in the Bode diagram of Fig. 5(c). The accuracy of the stability analysis is thus verified.

Since the system is unstable, the sensitivity analysis is further conducted. Both $\lambda_1(s)$ and $\lambda_2(s)$ are close to (−1, 0), thus they are the critical eigenloci. The Bode diagrams of the eigenloci in Fig. 5(c) also indicate that the critical frequency range should be around 630 Hz.

The port-level sensitivity analysis is conducted first. The port participation functions of $\lambda_1(s)$ and $\lambda_2(s)$ are calculated and their magnitude responses are compared in Fig. 7. It can be seen that in the critical frequency range, as marked in the red box, the sensitivity functions with respect to Ports 10-12 have larger magnitudes than others. These ports are associated with the Onshore Station 2. It is thus implied that the instability is more sensitivity to the Onshore Station 2.

The station impedance-level sensitivity is further analyzed. It is assumed that each MMC station has the same control for the positive pole and negative pole, thus the positive-pole and negative-pole impedances are the same. The normalized sensitivity functions of $\lambda_1(s)$ and $\lambda_2(s)$ are only calculated with respect to the positive-pole impedance of each station, whose magnitudes are compared in Fig. 8. It is seen that in the critical frequency range, as marked in the red box, the Onshore Station 2 has the largest sensitivity magnitude. It is implied that the instability is most sensitive to the Onshore Station 2, which agrees with the port-level sensitivity analysis.

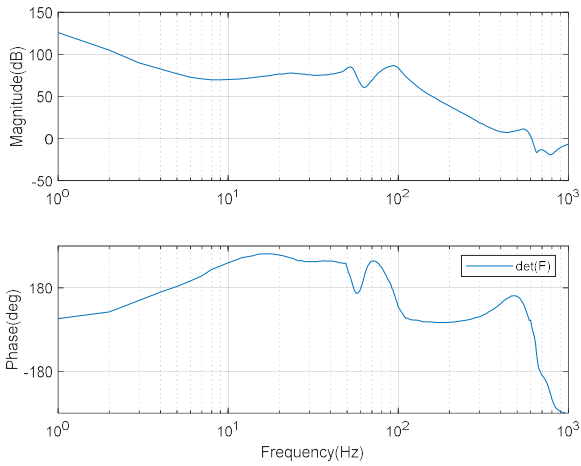

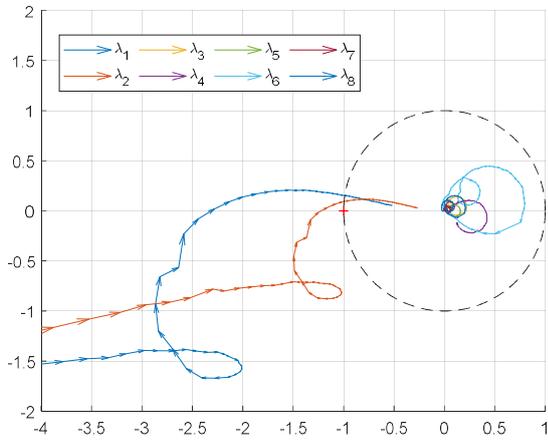

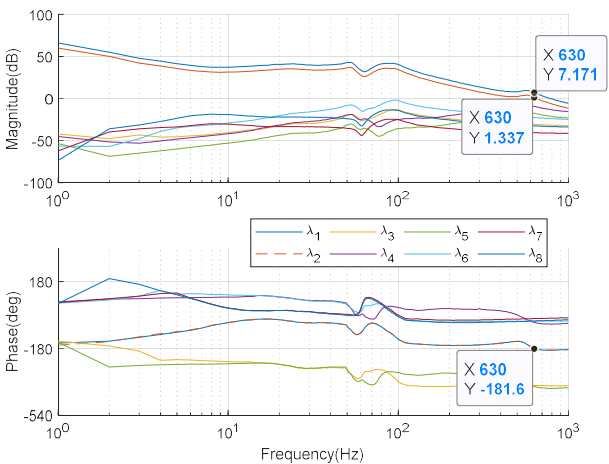

Fig. 5 Stability analysis for Case I. (a) Bode diagram of det($\mathbf{F}$(s)); (b) Nyquist diagrams of $\lambda_i$(s); (c) Bode diagrams of $\lambda_i$(s).

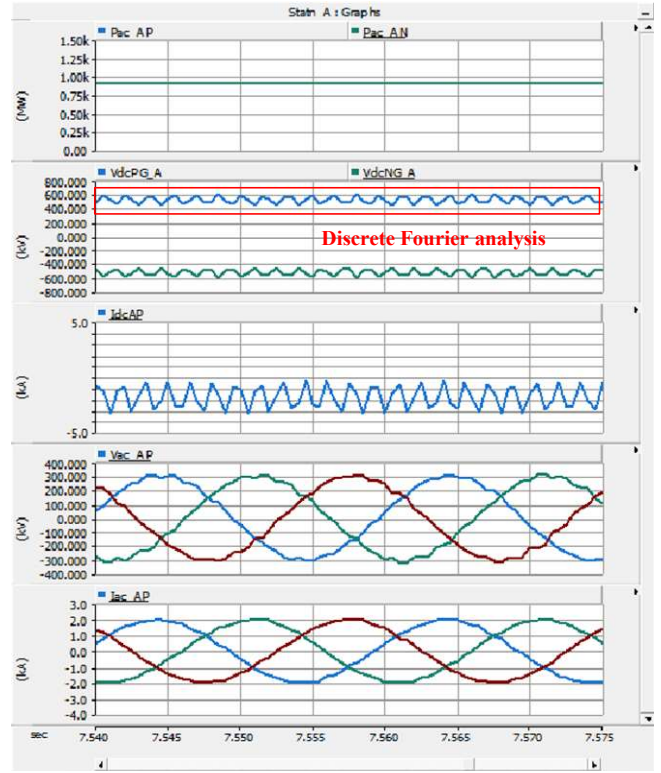

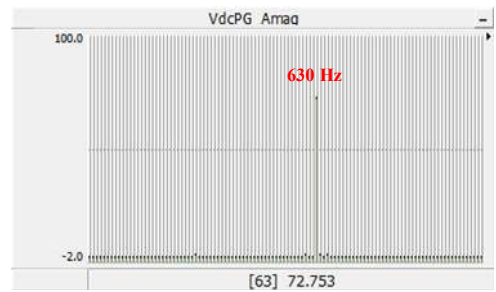

Fig. 6 EMT simulation results for Case I. (a) Time-domain waveforms; (b) Discrete Fourier analysis of DC-link voltage.

### B. Case II – Critically Unstable Case

The sensitivity analysis of Case I has indicated that Onshore Station 2 affects the stability most significantly. Thus, Case II is further analyzed by simply tuning the control parameters of the Onshore Station 2.

The stability analysis is provided in Fig. 9. It can be seen that now only $\lambda_1$ critically encircles (−1, 0), which implies the stability has been improved compared with Fig. 5 by changing the control parameters of the Onshore Station 2. This verifies the sensitivity analysis of Case I. The EMT simulation results are provided in Fig. 10, where almost none oscillation is observed, since the system is under a critically unstable condition. The sensitivity analyses at the port level and station impedance level are shown in Fig. 11 and Fig. 12, which also show similar results to Case I, indicating that the stability is more sensitive to the Onshore Station 2.

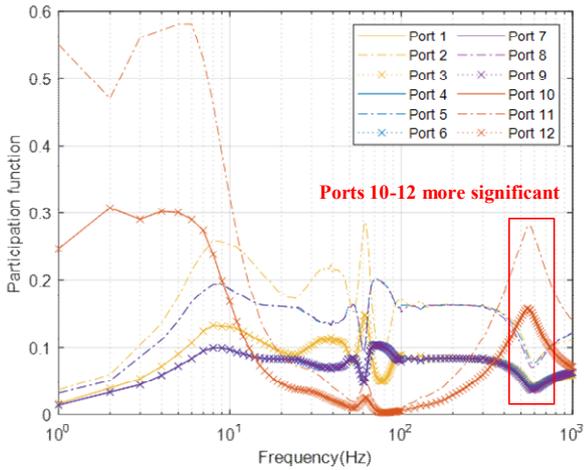

(a)

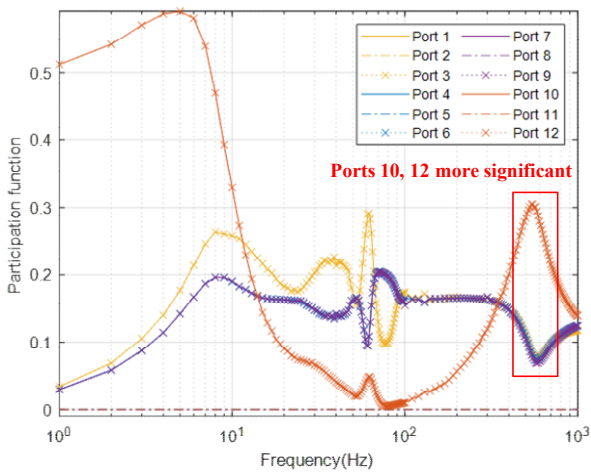

(b)

Fig. 7 Port participation analysis of $\lambda_1$ and $\lambda_2$ for Case I. (a) Participation analysis of $\lambda_1$; (b) Participation analysis of $\lambda_2$.

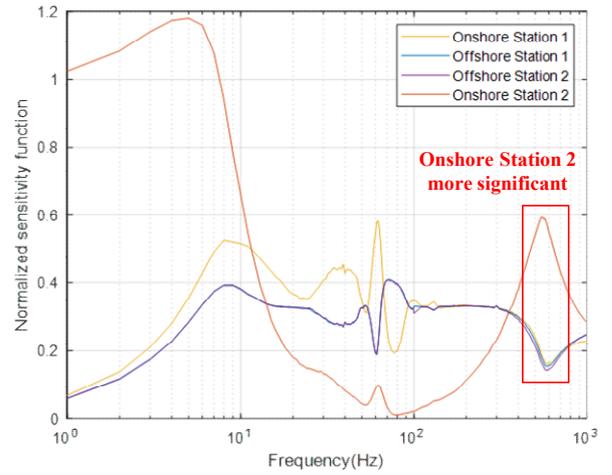

(a)

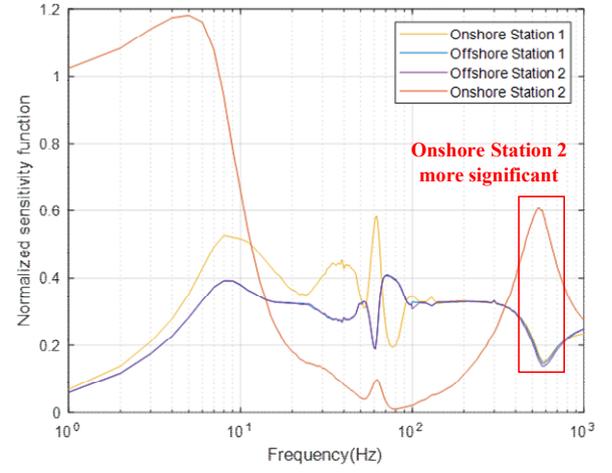

(b)

Fig. 8 Station impedance-level sensitivity analysis of $\lambda_1$ and $\lambda_2$ for Case I. (a) Sensitivity analysis of $\lambda_1$; (b) Sensitivity analysis of $\lambda_2$.

V. CONCLUSION

A multi-level sensitivity analysis approach has been developed to analyze the stability of multi-vendor, multi-terminal HVDC systems. This approach is derived based on frequency-domain sensitivity theory, which is thus readily applicable for black-box systems with only impedance representations.

The proposed approach allows for sensitivity analysis from the system level to converter control level, which provides insights to the system stability and design. More specifically,

- Port-level and station impedance-level sensitivity can help TSOs perform the system-level stability analysis and identify the root cause of potential instability.

- The system-level black-box sensitivity functions can also be transferred down to the converter manufacturers to help them further conduct the control-level sensitivity analysis and controller design for stability enhancement.

ACKNOWLEDGMENT

The research presented in this paper is financed by TenneT TSO GmbH.

REFERENCES

[1] C. Buchhagen, C. Rauscher, A. Menze and J. Jung, "BorWin1 - First Experiences with harmonic interactions in converter dominated grids," *International ETG Congress*; Die Energiewende - Blueprints for the new energy age, 2015, pp. 1-7.
[2] H. Saad, Y. Fillion, S. Deschanvres, Y. Vernay and S. Dennetière, "On Resonances and Harmonics in HVDC-MMC Station Connected to AC Grid," *IEEE Trans. Power Del.*, vol. 32, no. 3, pp. 1565-1573, June 2017.

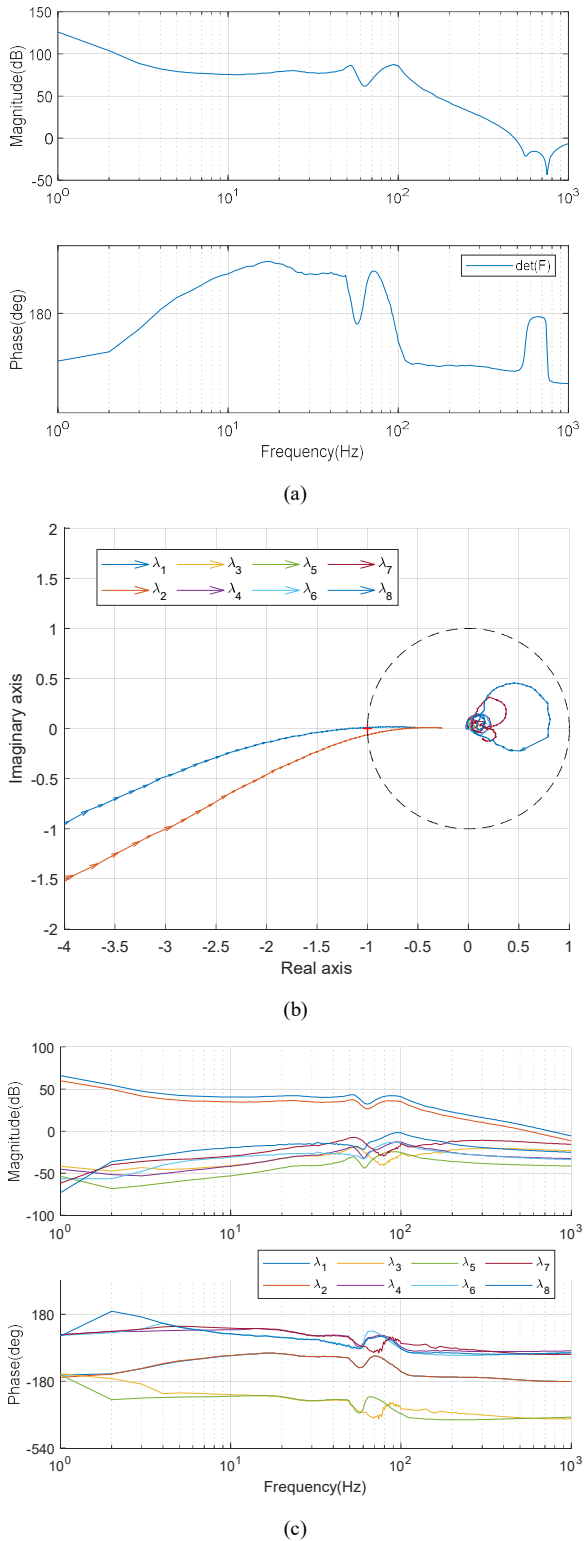

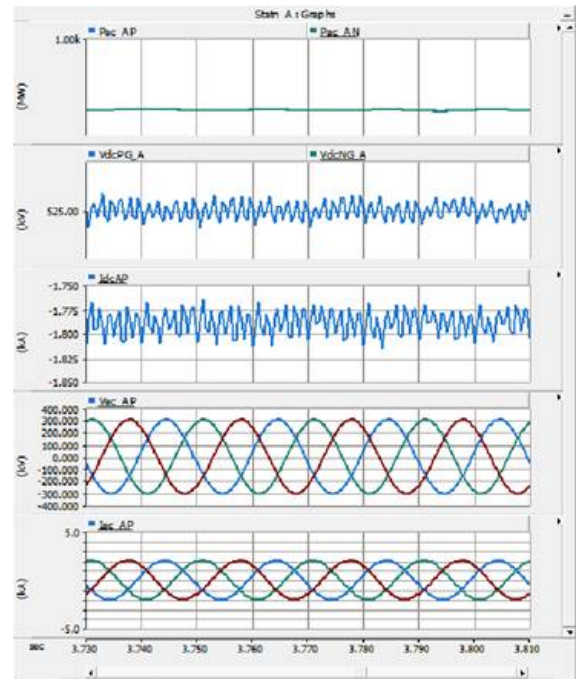

Fig. 10 EMT simulation results for Case II.

Fig. 9 Stability analysis for Case II. (a) Bode diagram of det(**F**(*s*)); (b) Nyquist diagrams of $\lambda_i(s)$; (c) Bode diagrams of $\lambda_i(s)$.

[3] ENTSO-E, "Position on Offshore Development – Interoperability," Jan. 2021. [Online]. Available: https://eepublicdownloads.entsoe.eu/clean-documents/Publications/Position%20papers%20and%20reports/210125_Offshore%20Development_Interoperability.pdf
[4] ENTSO-E, T&D Europe, WindEurope, "Workstream for the development of multi-vendor HVDC systems and other power electronics interfaced devices," Mar. 2021. [Online]. Available: https://eepublicdownloads.azureedge.net/clean-documents/RDC%20documents/210505%20Multi-Vendor-HVDC-workstream.pdf
[5] S. Liu, Z. Xu, W. Hua, G. Tang and Y. Xue, "Electromechanical Transient Modeling of Modular Multilevel Converter Based Multi-Terminal HVDC Systems," *IEEE Trans. Power Syst.*, vol. 29, no. 1, pp. 72-83, Jan. 2014.
[6] J. Lyu, X. Cai and M. Molinas, "Frequency Domain Stability Analysis of MMC-Based HVdc for Wind Farm Integration," *IEEE J. Emerging Sel. Topics Power Electron.*, vol. 4, no. 1, pp. 141-151, March 2016.
[7] G. Pinares and M. Bongiorno, "Modeling and Analysis of VSC-Based HVDC Systems for DC Network Stability Studies," *IEEE Trans. Power Del.*, vol. 31, no. 2, pp. 848-856, April 2016.
[8] T. Li, A. M. Gole and C. Zhao, "Harmonic Instability in MMC-HVDC Converters Resulting From Internal Dynamics," *IEEE Trans. Power Del.*, vol. 31, no. 4, pp. 1738-1747, Aug. 2016.
[9] H. Wu, X. Wang and Ł. H. Kocewiak, "Impedance-Based Stability Analysis of Voltage-Controlled MMCs Feeding Linear AC Systems," *IEEE J. Emerging Sel. Topics Power Electron.*, vol. 8, no. 4, pp. 4060-4074, Dec. 2020.
[10] X. Wang and F. Blaabjerg, "Harmonic Stability in Power Electronic-Based Power Systems: Concept, Modeling, and Analysis," *IEEE Trans. Smart Grid*, vol. 10, no. 3, pp. 2858-2870, May 2019.
[11] Y. Zhan, X. Xie and Y. Wang, "Impedance Network Model Based Modal Observability and Controllability Analysis for Renewable Integrated Power Systems," *IEEE Trans. Power Del.*, vol. 36, no. 4, pp. 2025-2034, Aug. 2021.
[12] Y. Zhu, Y. Gu, Y. Li and T. Green, "Participation Analysis in Impedance Models: The Grey-Box Approach for Power System Stability," *IEEE Trans. Power Syst.*, Early Access, doi: 10.1109/TPWRS.2021.3088345.

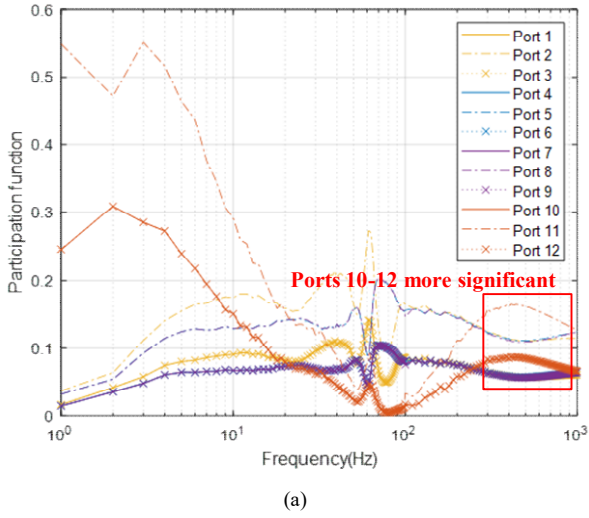

(a)

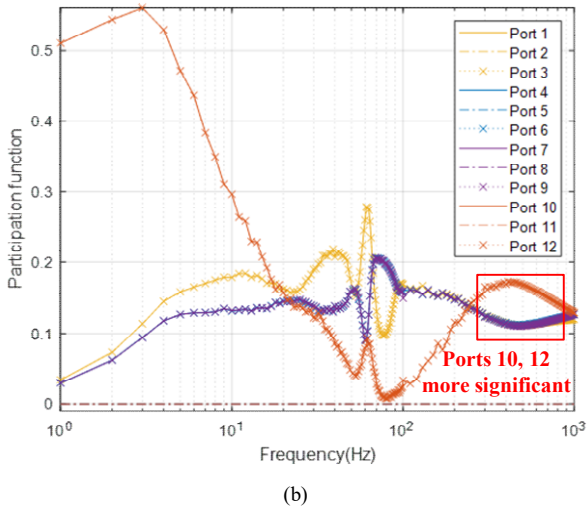

(b)

Fig. 11 Port participation analysis of $\lambda_1$ and $\lambda_2$ for Case II. (a) Participation analysis of $\lambda_1$; (b) Participation analysis of $\lambda_2$.

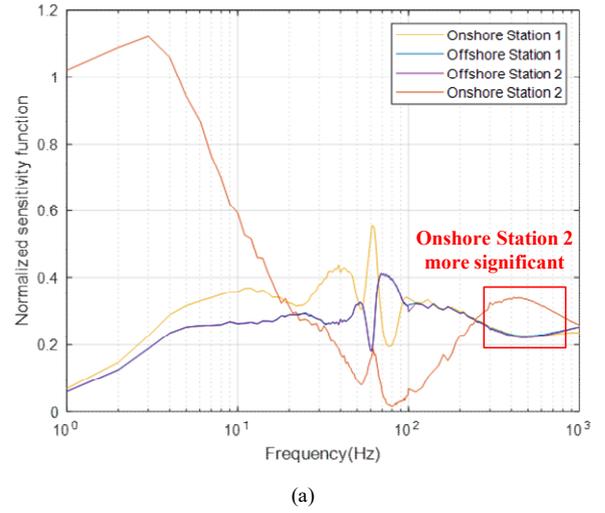

(a)

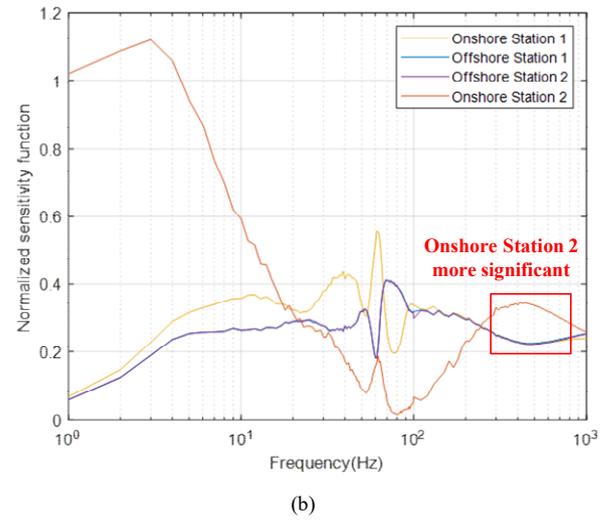

(b)

Fig. 12 Station impedance-level sensitivity analysis of $\lambda_1$ and $\lambda_2$ for Case II. (a) Sensitivity analysis of $\lambda_1$; (b) Sensitivity analysis of $\lambda_2$.

[13] H. Zhang, X. Wang, M. Mehrabankhomartash, M. G. Saeedifard, Y. Meng and X. Wang, "Harmonic Stability Assessment of Multi-terminal DC (MTDC) Systems Based on the Hybrid AC/DC Admittance Model and Determinant-based GNC," *IEEE Trans. Power Electron.*, Early Access, doi: 10.1109/TPEL.2021.3103797.
[14] H. Wu, X. Wang, Y. Liao, M. Ndreko, R. Dimitrovski, and W. Winter, "Development of the toolbox for AC/DC impedance matrix measurement of MTDC system," *in Proc. 20th Wind Integration Workshop*, Sept. 2021, Berlin, Germany.
[15] A. MacFarlane, "Return-difference and return-ratio matrices and their use in analysis and design of multivariable feedback control systems," in *Proc. Institution Elect. Eng.*, vol. 117, no. 10, pp. 2037–2049, 1970.
[16] P. M. Frank, *Introduction to System Sensitivity Theory*, New York: Academic Press, 1978.
[17] P. Kundur, *Power system stability and control*. McGraw-hill New York, 1994.
[18] Y. Liao and X. Wang, "Impedance-Based Stability Analysis for Interconnected Converter Systems With Open-Loop RHP Poles," *IEEE Trans. Power Electron.*, vol. 35, no. 4, pp. 4388-4397, Apr. 2020.